\documentclass{kluwer}    
\usepackage{graphicx} 

\newdisplay{guess}{Conjecture}

\begin{document}
\begin{article}
\def\gtrsim{\mathrel{\hbox{\rlap{\hbox{\lower4pt\hbox{$\sim$}}}\hbox{$>$}}}}
\def\lesssim{\mathrel{\hbox{\rlap{\hbox{\lower4pt\hbox{$\sim$}}}\hbox{$<$}}}}
\begin{opening}         
\title{The Stellar IMF in Very Metal-Deficient Gas} 
\author{Fumitaka \surname{Nakamura}$^*$ and Masayuki \surname{Umemura}$^{**}$}  
\runningauthor{F. Nakamura \& M. Umemura}
\runningtitle{IMF in Metal-Deficient Gas}
\institute{$^*$Niigata University and UC Berkeley,
 $^{**}$University of Tsukuba}
\begin{abstract}
In the context of the star formation through the fragmentation of
an extremely metal-deficient protogalactic cloud,
the gravitational collapse of filamentary gas clouds is explored with
H$_2$ and HD chemistry.
It is found by 1D hydrodynamical simulations 
that the cloud evolution is prescribed mainly by
the initial density ($n_0$) and H$_2$ abundance ($x_{\rm H_2,0}$).
In particular, it turns out that the evolution of low-density filaments 
($n_0 \lesssim 10^5$ cm$^{-3}$) bifurcates at a critical H$_2$ abundance
 of $x_{\rm H_2,cr}\simeq 3\times 10^{-3}$,
 beyond which HD cooling overwhelms H$_2$ cooling.
The numerical results indicate that the stellar IMF is likely to be 
double-peaked and deficient in sub-solar mass stars, where 
the high mass peak of the IMF is around $10M_\odot$ or $10^2M_\odot$,
dependently on the initial density and H$_2$ abundance.
If the gas in protogalactic clouds is photoionized by UV radiation or
shock-heated, the H$_2$ abundance could exceed
 $x_{\rm H_2,cr}\simeq 3\times 10^{-3}$
by H$^-$ reactions. Then, the high mass peak would be $O(10) M_\odot$.
\end{abstract}
\end{opening}           

\vspace*{-0.5cm}
\section{Introduction}  
\vspace*{-0.3cm}

Star formation in the early universe should have proceeded
 in very metal-deficient gas.
When the metallicity is lower than $Z \lesssim 10^{-2}Z_\odot$, 
the thermal property of such metal-deficient gas
 is essentially the same as that of the metal-free gas because  
 cooling by heavy metals is less effective than primordial molecular cooling.
Among the primordial molecules, the importance of H$_2$
 on the star formation in the early universe have been emphasized
 by many authors because they are the most abundant molecules
 in metal-free gas.
H$_2$ molecules provide significant cooling through rotational and
 vibrational transitions 
 which can lower the gas temperature down to a few hundred K
 and accordingly reduce the Jeans masses to stellar mass scales.

HD molecules are considered to be the second
 most abundant molecules in primordial gas 
during the post-recombination era.
In spite of its low abundance 
([HD/H$_2$] $\sim 10^{-3}$), HD can provide significant cooling at low 
temperature gas because HD has higher radiative transition
 probabilities and lower excitation temperatures than H$_2$.
Therefore, HD cooling can lower the gas temperature down to 
$T\lesssim$ 100K and accordingly, the Jeans mass is also reduced.  

In this paper, we  examine the effects of the HD cooling
on the formation of stars in very metal-deficient gas, and 
elucidate the role of HD molecules for the stellar initial mass function
(IMF) there.

\clearpage
\section{Numerical Model and Methods}
\vspace*{-0.3cm}

Our numerical model and method are the same as
those of \citeauthor{Nakamura01a} (\citeyear{Nakamura01a}).
We consider an infinitely long cylindrical gas cloud. 
We deal with the following 14 species:
e, H, H$^+$, H$^-$, H$_2$, H$_2^+$, He, He$^+$, He$^{++}$,
D, D$^+$, D$^-$, HD, and HD$^+$.

The density is assumed to be uniform along the cylinder axis 
and the radial distribution is expressed as 
$\rho =\rho _0 \left(1+r^2/R_{0}^2\right)^{-2}$, 
where $R_0=\sqrt{2fkT_0/(\pi G\rho _0\mu)}$ is the effective radius,
$\rho_0$ is the central mass density, $T_0$ is the initial
gas temperature, $\mu$ is the mean molecular weight, 
and $f$ is the ratio of the gravitational force
to the pressure force.
Our model is specified by four parameters: $n_0$,
$T_0$, $f$, and the H$_2$ abundance $x_{\rm H_2, 0}$.
It should be noted that a higher $x_{\rm H_2,0}$ can be
translated to a higher electron abundance of a parent cloud.
This is because when the ionization degree of the parent cloud
is higher, the H$_2$ molecules form more efficiently via radiative 
reactions of $\rm H+e \rightarrow H^- +{\it h}\nu$ and 
$\rm H+H^- \rightarrow H_2 + e$.

We take into account the following thermal processes: 
(1) H cooling by radiative recombination, 
collisional ionization, and collisional excitation, 
(2) H$_2$ line cooling, 
(3) cooling by H$_2$ collisional dissociation, 
(4) heating by H$_2$ formation, and 
(5) HD line cooling.

\vspace*{-0.5cm}
\section{Numerical Results}
\vspace*{-0.3cm}

In this section, we examine the collapse of the filaments, 
taking into account the H$_2$ and HD cooling.
As shown below, there is a critical initial H$_2$ abundance,
above which HD cooling predominantly regulates the cloud evolution. 
Also, it is found that the HD cooling does not play an important role
 for high-density gas.
Thus, the evolution of the primordial filaments is classified
into three cases, depending upon the initial density and
initial H$_2$ abundance; 
(1) low-density filaments with high H$_2$ abundance, (2) low-density
filaments with low H$_2$ abundance, and (3) high-density filaments.

For low density filaments with $x_{\rm H_2, 0}\gtrsim 3\times 10^{-3}$, 
 HD cooling overwhelms H$_2$ cooling during the
contraction, and the temperature descends down to $T\sim 50$ K accordingly. 
When the density reaches a critical density of HD 
($n_{\rm cr} \sim 10^{4-5}$ cm$^{-3}$),
the cloud contraction tends to become quasistatic.
Thus, the fragmentation will take place after that stage.
For low density filaments with $x_{\rm H_2, 0}\lesssim 3\times 10^{-3}$, 
because of the low $x_{\rm H_2}$, the
temperature stays a relatively high value at a few hundred K, and
HD cooling does not play a role during the contraction.
The contraction proceeds quasi-statically
after the density reaches the critical density of H$_2$
($n>n_{\rm H_2, cr}=10^{3-4}$ cm$^{-3}$).
As for high density filaments with $n_0 \gtrsim 10^5$ cm$^{-3}$,
the temperature stays at a relatively high value 
 because both H$_2$ and HD cooling is saturated,
and therefore HD does not play a significant role for the thermal evolution.
The contraction of such a high-density filament is accelerated
 by the effective three-body H$_2$ formation when the density reaches
 $10^{8-9}$ cm$^{-3}$.
Then, the fragmentation is not expected to take place until the cloud
becomes opaque to H$_2$ lines at the density of $10^{12-13}$ cm$^{-3}$.

%
\begin{figure}
\centering
\includegraphics[width=.9\textwidth]{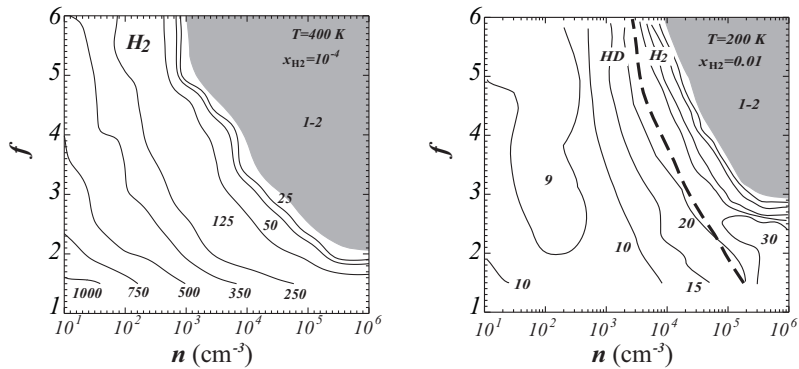}
\caption[]{
Mass distributions of the fragment 
derived from 1D simulations
for the models with (a) $x_{\rm H_2, 0} = 1\times 10^{-4}$ and 
 (b) $1\times 10^{-2}$, respectively.
The abscissa and ordinate denote the initial central density 
and the parameter $f$, respectively.
The solid lines denote the contours of the fragment mass which are
labeled with adjacent numbers.
The dashed lines show the lines at which the HD cooling rate
is equal to H$_2$ cooling rate at the epoch of fragmentation.
In the left regions of the dashed lines, HD cooling is more 
efficient than H$_2$ cooling.
}
\label{fig:1}
\end{figure}

Figures \ref{fig:1}a and \ref{fig:1}b
show the distributions of the fragment mass
derived from the 1D simulations
 (see \citeauthor{Nakamura01b} (\citeyear{Nakamura01b}) for more detail).
For the models with $x_{\rm H_2, 0} < x_{\rm H_2, cr} \approx 
3\times 10^{-3}$ (Figures \ref{fig:1}a), 
the mass distribution of the fragments is quite similar to the case
without HD (Paper II), because HD cooling does not play an
important role in the thermal evolution of the filaments.
There is a steep boundary at $n\sim 10^4 - 10^5$ cm$^{-3}$ 
in the mass distribution of the fragments for $f\gtrsim 3$. 
For the models with $n_0 \gtrsim 10^5$ cm$^{-3}$, 
the fragment masses take their minima at $1\sim 2$ M$_\odot$,
 whereas, for the models with $n_0 \lesssim 10^5$ cm$^{-3}$,
 they are greater than $\sim 10^2$ M$_\odot$.
[For the models with low densities ($n_0\lesssim 10^5$ cm$^{-3}$), 
 the contraction becomes quasistatic when the density reaches the
 critical density of H$_2$ and then linear density fluctuations can 
 grow nonlinearly before the three-body H$_2$ formation becomes
 dominant ($n_0\gtrsim 10^{8-9}$ cm$^{-3}$).
In contrast, for the models with high densities
 ($n_0\gtrsim 10^{5-6}$ cm$^{-3}$),
the contraction time does not exceed the fragmentation time
 until the H$_2$ lines become optically thick
at $n\sim 10^{12}-10^{13}$ cm$^{-3}$.]
This sensitivity in the fragment mass comes from 
the rapid increase in $x_{\rm H_2}$ due to
 the three-body reactions.

On the other hand, when $x_{\rm H_2,0} > 3\times 10^{-3}$, 
HD cooling is more effective than H$_2$ cooling for low-density filaments.
Thus, the maximum mass of low-density region reduces to
a few tens M$_\odot$. 
The minimum mass does not change because HD is not
a dominant coolant in the evolution of dense filaments.
Therefore, similarly to the models with low $x_{\rm H_2,0}$
the dependence of the fragment mass on $n_0$
 exhibits a steep boundary
 around $n_0 = 10^4 - 10^5$ cm$^{-3}$ for $f\gtrsim 3$.

The existence of a steep boundary in the mass distribution 
implies that the IMF is
 likely to be bimodal if both low-density and high-density filaments
bear stars.
The low-mass peak is around a few $M_\odot$, which is not sensitive to
the abundance of H$_2$ formed in a parent cloud.
The high-mass peak is $\approx 10^2 M_\odot$ if $x_{\rm H_2,
0}\lesssim 3\times 10^{-3}$, while it is
$\approx 10 M_\odot$ if $x_{\rm H_2, 0} \gtrsim 3\times 10^{-3}$.
In the next section, we discuss some implications of
such bimodal IMF in the very metal-deficient gas.

\vspace*{-0.6cm}
\section{Implications for Galaxy Formation}
\vspace*{-0.3cm}

Recently, \citeauthor{Susa00} (\citeyear{Susa00}) 
investigated the pancake collapse
 of pregalactic clouds under UV background radiation.
They found that once the pancaking disk is shielded against
 external UV radiation, the H$_2$ molecules form efficiently 
via H$^-$ reaction with abundant free electrons produced by
 UV background, and the resultant abundance reaches 
$x_{\rm H_2} \approx 3\times 10^{-3}$.
The pancake disks probably fragment into filaments in which 
stars would form.
In this case, HD cooling is expected to become efficient
 in low-density filaments and then, 
 the high mass peak of the IMF would go down to
 $\sim$ 10 M$_\odot$.

The time-decreasing UV background radiation is also likely to 
 influence star formation in dwarf galaxies
 which may be related to excess number count 
 of faint blue objects observed in the Hubble Deep Field.  
\citeauthor{Corbelli97} (\citeyear{Corbelli97}) studied the effects of
 the declined UV background
 on the thermal evolution of the protogalaxies.  
They found that there is a critical redshift
 of $z \sim 1-2$, below which  
 the declined UV radiation is shielded by the gas
 disks where the H$_2$ abundance reaches $10^{-2}$ owing to high
 ionization degree by the UV radiation.
The enhanced H$_2$ formation promotes a rapid
 transition toward the cold H I phase with $\sim 10^2$ K.
Thus, in such dwarf galaxies, high mass peak of the IMF 
 would decrease to $\sim$ 10 M$_\odot$ owing to HD cooling.


\theendnotes

\end{article}
\end{document}